\documentclass[12pt]{article} 
\usepackage{latexsym,epsf,epsfig,wrapfig}
\begin{document}

\title{\bf Nuclear effects in positive 
pion electroproduction on the deuteron 
near threshold}  
%\vspace{5mm}
\author{L. Levchuk$^1$, L. Canton$^2$, and A. Shebeko$^1$} 
\date{}
\maketitle
\begin{center}
{\em 
$^1$ Kharkov Institute of Physics and Technology, 61108 Kharkov, Ukraine,\\
$^2$ Sezione di Padova, Istituto Nazionale di Fisica Nucleare, I-35131 Padova, Italy 
}
\end{center}

\vspace{5mm}
%\begin{center}
%ABSTRACT

%\vspace{5mm}
%\begin{minipage}{130 mm}
%\small
\begin{abstract}
Positive pion electroproduction from the deuteron near threshold has been 
considered within an approach based on the unitary transformation method. The 
gauge independence of the treatment is provided by using an explicitly gauge 
independent expression for the reaction amplitude. The results of calculations 
for kinematics of the experiments on forward-angle {$\pi ^+ $} meson 
electroproduction accomplished at Saclay and Jefferson Laboratory 
are discussed and compared with those given by the impulse 
approximation. It is shown that the observed behaviour of the cross sections is 
in accordance with the calculations based on the pion-nucleon dynamics. 
In particular, the pion production rate suppression 
in the $^2{\rm H}({\rm e,e}'\pi ^+$)nn reaction compared to that for the 
$^1{\rm H}({\rm e,e}'\pi ^+$)n one can be due to such ``nuclear medium'' effects 
as nucleon motion and binding along with Pauli blocking in the final 
nn state.

\end{abstract}
%\end{minipage}
%\end{center}

\newcommand{\ve }[1]{{\mbox{\boldmath $#1$}}} 
\newcommand{\vsm }[1]{{\mbox{\footnotesize{\boldmath $#1$}}}} 
\thispagestyle{empty}

%\bigskip
%\centerline{\bf I. INTRODUCTION}
%\bigskip

\section{Introduction}

The experimental facilities (such as those at Jefferson Laboratory) put into operation 
for the last decade have opened new horizons in studying the 
structure of nucleons and nuclei. A lot of valuable information can be obtained, 
%%%%%%%%%%%%%%%%%%%%%%%% FIGURE 1 %%%%%%%%%%%%%%%%%%%%%%%%%%%%%%%%%%%%%%%%%%%%%% 
%%\begin{wrapfigure}[17]{r}{9cm}
%%\epsfig{figure=figure1.eps,width=9cm}
%%{\small FIG. 1. Pion electroproduction on the deuteron: a) kinematics and 
%%b) the IA ansatz.}
%%\end{wrapfigure}
%\begin{figure}[h]
%\epsfxsize=10cm
%\epsffile{figure1.eps}
%
%{\small FIG. 1. Pion electroproduction on the deuteron: a) kinematics and 
%b) the IA ansatz.}
%\end{figure}
%%%%%%%%%%%%%%%%%%%%%%%%%%%%%%%%%%%%%%%%%%%%%%%%%%%%%%%%%%%%%%%%%%%%%%%%%%%%%%%% 
\begin{figure}[hbtp]
  \begin{center}
    \resizebox{10cm}{!}{\includegraphics{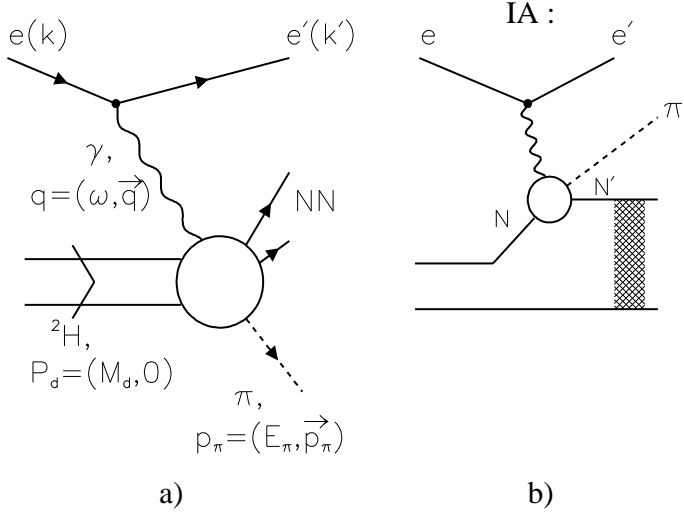}}
    \caption{Pion electroproduction on the deuteron: a) kinematics and b) the IA ansatz.}
    \label{1}
  \end{center}
\end{figure}
in particular, in high precision experiments on pion electroproduction, including 
measurements of polarization observables and the separation of the structure functions (SF's). 
These possibilities have already triggered a number of novel theoretical approaches (see, 
e.g., Ref.~\cite{SurGro}) developed for the treatment of pion photo(electro-)production off the 
nucleon. Furthermore, the data~\cite{MazBeck} on neutral pion production on the proton near threshold 
also renewed the interest in this reaction. Applications~\cite{Meissner} of the chiral perturbation 
theory to the threshold pion photo- and electroproduction resulted in new conclusions about 
the predictive force of the low-energy theorems, which, in turn, have been rederived~\cite{NausKoch}.

More understanding of mechanisms of pion photo(electro-)production can be gained in 
studies of the reaction on bound systems of nucleons. Apart from offering means to 
extract the amplitude of pion production off the neutron, the processes like (see Fig.~\ref{1}a) 
$$
\gamma\ (\gamma ^* ) + ^2{\rm H} \rightarrow \pi + {\rm NN} \eqno(1)
$$
could shed light on the details of the nuclear structure.

For the description of photomeson processes on nuclei, one usually addresses the impulse 
approximation (IA)~\cite{GoldWats} 
(see Fig.~\ref{1}b). Within this approach, the two-body mechanisms of the reaction (1) are 
disregarded along with binding effects in the one-body contribution. The latter assumes 
the validity of the on-energy-shell condition
$$
E_{\pi} + E_{\rm N'} - \omega - E_{\rm N} = 0 \eqno(2)
$$
for the elementary $\gamma {\rm N}\rightarrow \pi {\rm N}'$ amplitude (the notations 
are given in Fig.~\ref{1}). This condition helps one to provide the gauge independence (GI)  
of calculations within the IA. However, Eq.~(2) may contradict the energy 
conservation law for the reaction (see, e.g., the discussion in Ref.~\cite{LS95}). In general, 
there are no enough grounds to use the IA in studies of the reaction, when, in  
particular, high-momentum components of nuclear wave functions are probed. 
Moreover, for exclusive and semiexclusive reactions like $^2{\rm H}(\gamma ,\pi )$NN 
and $^2{\rm H}({\rm e,e}'\pi )$NN with the given pion 3-momentum, Eq.~(2) may overdetermine 
the reaction kinematics providing a different value for this vector. This leads to 
violation of the transformation properties of observables for these reactions in calculations 
based on the IA (see Ref.~\cite{LS99}). 
 
We present in this paper some applications of an alternative formalism 
developed in Refs.~\cite{LS95,LS89} 
for the description of many-body currents and off-shell effects in the theory of 
photomeson processes on nuclei. Our approach is based upon the unitary 
transformation method~\cite{Okubo} in combination with the idea~\cite{Foldy,FriarFall,LS93} to provide the GI 
of calculations through an extension of the Siegert theorem.

The formalism employed in our consideration is outlined briefly in Sec.~2. 
The practical recipe for providing the GI of calculations is given in Sec.~3. 
In Sec.~4, our results for the cross section of the $^2{\rm H}({\rm e,e}'\pi^+ )$nn 
reaction are presented and discussed. Sec.~5 contains concluding remarks.

%\bigskip
%\centerline{\bf II. UNDERLYING FORMALISM}
%\bigskip
%\bigskip
\section{Underlying formalism}
%\bigskip

The present consideration refers to the d(e,e$'\pi ^{+}$)nn reaction. Its 
amplitude to the one-photon exchange approximation (OPEA) is given by
$$
T_{if}=[2(2\pi )^3 \omega ]^{-{{1}\over{2}}}\ {\varepsilon } _{\nu}(\ve{q})\ J_{if}^{\nu } (\ve{q})\ , 
\eqno(3)
$$
$$
J_{if}^{\nu } (\ve{q}) \equiv \langle\pi ^{+}{\rm nn;out}\mid J^{\nu}(0) \mid {\rm d}\ \rangle\ ,
$$
where ${\varepsilon}_{\nu}(\ve{q})$ is the (virtual) photon polarization vector 
[$q\cdot\varepsilon (\ve{q})=\omega \varepsilon_0(\ve{q})-\ve{q\varepsilon} (\ve{q})=0$], 
$q=(\omega ,\ve{q})$ is the 4-momentum transfer (the photon momentum 
[see Fig.~\ref{1}a]), and $J^{\nu}(0)$ is the electromagnetic (e.m.) current density at the 
space-time point $x=(t,\ve{x})=0$. The corresponding GI requirement reads:
$$
q_{\mu }\ J_{if}^{\mu } (\ve{q}) =0\ . 
\eqno(4)
$$
Using the reduction technique prescriptions (see, e.g., Ref.~\cite{LSZ}), one has
$$
\langle {\pi}^+ {\rm nn ;out}\mid J_{\nu}(0)\mid {\rm d}\rangle =
(2(2\pi )^3 E_{\pi})^{-{ 1\over 2}}\ \langle {\rm nn;out}\mid T_{\nu} 
(\ve{p}_{\pi})\mid{\rm d}\rangle\ ,\eqno(5)
$$
$$
T_{\nu} (\ve{p}_{\pi}) = i\int {\rm e}^{ip_{\pi} x}
(\Box + m_{\pi}^2 )\ T\{{\phi} (x) J_{\nu}(0)\}\ {\rm d}^4 x\ ,
$$
where ${\phi}(x)$ is the pion field operator.

Then, following Refs.~\cite{LS95,LS89}, we apply the unitary transformation method~\cite{Okubo} to construct 
the effective operator acting in the space of states $\mid \chi _{\rm d} \rangle 
\ (\mid \chi _{\rm nn} \rangle )$ with no mesons:
$$
\langle {\rm nn;out}\mid T_{\nu} (\ve{p}_{\pi})\mid
{\rm d}\rangle = \langle {\chi}_{\rm nn}^{(-)} \mid [T_{\nu} (\ve{p}_{\pi}) ]^{\rm eff} 
\mid {\chi}_{\rm d} \rangle\ .\eqno(6)
$$
This operator consists of the one-body contribution $T^{[1]}$ and the contribution 
$T^{[2]}$ due to pion production by a couple of nucleons:
$$
[T_{\nu} (\ve{p}_{\pi})]^{\rm eff} = T_{\nu}^{[1]}+T_{\nu}^{[2]}\ .\eqno(7)
$$
{\it A priori}, one cannot exclude the mechanisms such that the photon is 
absorbed by one nucleon, and the pion is emitted by another one. However, the 
perturbation analysis~\cite{LS89} has shown that the corresponding contribution to the 
lowest degree in the $\pi$NN coupling constant $g$ is dropped out, and the perturbation 
series for $T^{[2]}$ begins with terms of order of $g^3$. 
This cancellation of the simplest two-body pion production contributions is similar 
to the so-called Jennings cancellation mechanism~\cite{Jennings} in $\pi$d scattering.
To calculate the one-body 
contribution $T^{[1]}$, one may use an amplitude of pion production on a free nucleon 
obtained within some dynamical approach making in the corresponding expression 
the off-shell transition $q_{\mu} \rightarrow \widetilde{q} _{\mu} \equiv 
(E_{\rm N'} + E_{\pi} - E_{\rm N} ,\ \ve{q} )$ (see the discussion in Ref.~\cite{LS95}). This
transition reflects the nuclear medium influence on 
the one-body pion production mechanism. It implies, in particular, that the hadrons 
participating in pion production on a bound nucleon even by a real photon  ($q^2=0$) 
are not considered as point-like particles, since $\widetilde{q}^2 \neq 0$ in 
this case. It should be stressed that, unlike the IA (see Eq.~(2)), no extra  
kinematical assumptions are made, and the hadron momenta take on values 
determined by kinematics of the reaction on the bound system (see Fig.~\ref{1}a). 
Thus, the structure functions determining, e.g., the differential cross sections of 
pion electroproduction on the deuteron (see Sec.~4) preserve the properties given 
by their definition (see discussion in Ref.~\cite{LS99}). 

%\bigskip
%\centerline{\bf III. GAUGE INDEPENDENT EXPRESSION FOR THE AMPLITUDE}
%\bigskip
%\bigskip
\section{Gauge independent expression for the amplitude}
%\bigskip

In general, calculations based on the contribution $T^{[1]}$ only do not satisfy 
the GI requirement (4). Further difficulties can stem from the necessity to take into 
account the structure of hadrons, when describing pion electroproduction at $q^2\neq 0$. 
Then, an incompleteness of the description may lead to results, which are not gauge 
independent even in case of the reaction on a free nucleon (when the IA assumption (2) 
holds). To restore the GI of the treatment, one often adds an extra term to the amplitude 
making the subtraction 
$$
J_{\mu} \rightarrow J_{\mu}-q_{\mu}\ q\cdot J/q^2\ .\eqno(8)
$$
Of course, this procedure cannot reflect the complexity of the reaction mechanisms 
such as, e.g., the two-body processes. Moreover, it does not affect 
the transverse components of the transition matrix and is not unambiguous 
admitting extra subtraction of an arbitrary vector $X_{\mu}$ such that $q\cdot X = 0$.

In our consideration, to provide the GI of calculations, we make use of the
extension~\cite{Foldy,FriarFall,LS93} 
of the Siegert theorem expressing the amplitude in an explicitly gauge independent way 
through the Fourier transforms of electric ($\ve{E}(\ve{q})$) and magnetic ($\ve{H}(\ve{q})$) 
field strengths, 
$$
T_{if}= \ve{E}(\ve{q}) {\ve{D}}_{\rm if} + \ve{H}(\ve{q}) {\ve{M}}_{\rm if} \ ,
\eqno(9) 
$$
$$
\ve{E}(\ve{q})= i[2(2\pi )^3 \omega ]^{-{1\over 2}} \ (\omega \ve{\varepsilon} - 
{\varepsilon}_0 \ve{q}) \ ,
$$
$$
\ve{H}(\ve{q})= i[2(2\pi )^3 \omega ]^{-{1\over 2}} \ [\ve{q} \times \ve{\varepsilon} ] \ , 
$$
with $\ve{D}_{\rm if}$ and $\ve{M}_{\rm if}$ being matrix elements of 
generalized electric and magnetic dipole moments of the hadronic system containing 
the information on the nuclear dynamics. 

To get representation (9) (see Ref.~\cite{LS93}), consider expression 
$$
\delta (\ve{P}_i + \ve{q}- \ve{P}_f ) \langle \ve{P}_f , f \mid J^{\mu}(0) \mid \ve{P}_i , i \rangle 
= (2\pi )^{-3} \int {\rm exp} ( i \ve{q} \ve{s} ) j_{if}^{\mu} (\ve{s}) {\rm d}\ve{s} \ ,
\eqno(10)
$$
$$
j_{if}^{\mu} (\ve{s}) \equiv \left( \rho_{if} (\ve{s}) , \ve{j}_{if} (\ve{s}) \right)  
= \langle \ve{P}_f , f \mid J^{\mu}(\ve{s}) \mid \ve{P}_i , i \rangle = 
\langle \ve{P}_f , f \mid J^{\mu}(0) \mid \ve{P}_i , i \rangle \ 
{\rm e}^{-i(\vsm{P}_f - \vsm{P}_i ) \vsm{s} } \ .
$$
Multiplying the space part of matrix element (10) by an {\em arbitrary} 
vector $\ve{\varepsilon}(\ve{q})$ and applying the Foldy trick~\cite{Foldy} 
$$
\ve{\varepsilon} {\rm e}^{i\vsm{q}\vsm{s} } = \int_0^1 
\{
\nabla _{\ve{s}} ( \ve{\varepsilon} \ve{s} {\rm e}^{i \lambda \vsm{q}\vsm{s} } )\ -\ 
i \lambda \ve{s}\times [\ve{q}\times\ve{\varepsilon} ] {\rm e}^{i \lambda \vsm{q}\vsm{s} } 
\}
\ {\rm d}\lambda \ ,
$$
with help of the GI condition $ {\rm div}\ve{j}_{if}(\ve{s}) =-i(E_f -E_i)\rho_{if}(\ve{s})$, we get 
$$
\delta (\ve{P}_i + \ve{q} - \ve{P}_f ) \langle \ve{P}_f , f \mid \ve{J}(0) \mid \ve{P}_i , i \rangle = 
i (E_f - E_i ) \ve{d}_{if} (\ve{q}) - i [ \ve{q} \times \ve{m}_{if} (\ve{q}) ] \ , 
\eqno(11)
$$
$$
\ve{d}_{if} (\ve{q}) = (2\pi )^{-3} \int_0^1 {\rm d}\lambda \int {\rm e}^{i\lambda\vsm{q}\vsm{s} } \ve{s} 
\rho_{if} (\ve{s})\ {\rm d}\ve{s} \ ,
$$
$$
\ve{m}_{if} (\ve{q}) = (2\pi )^{-3} \int_0^1 \lambda{\rm d}\lambda \int {\rm e}^{i\lambda\vsm{q}\vsm{s} } 
\left[ \ve{s} \times \ve{j}_{if} (\ve{s})\right] \ {\rm d}\ve{s} \ .
$$
Then, due to charge conservation, one may write 
$$
\int i(E_f - E_i ) \ve{q} \ve{d}_{if} (\ve{q})\ {\rm d}\ve{P}_i  
= \omega \langle \ve{P}_f , f \mid J^0 (0) \mid \ve{P}_f - \ve{q}, i \rangle 
$$
$$
- (E_f - E_i (\ve{P}_f ) ) \langle \ve{P}_f , f \mid J^0 (0) \mid \ve{P}_f , i \rangle 
= 
\omega \langle \ve{P}_f , f \mid J^0 (0) \mid \ve{P}_f - \ve{q} , i \rangle 
\eqno(12)
$$
Integration of Eq.~(11) over $\ve{P}_i$ with taking into account relationship (12) 
results in representation (9), with quantities ${\ve{D}}_{if}$ and ${\ve{M}}_{if}$ 
being defined as 
$$
{\ve{D}}_{if}= - \int {{E_f - E_i}\over {}\omega } \ve{d}_{if} (\ve{q})\ {\rm d} \ve{P}_i \ ,
\ \ \ \ {\ve{M}}_{if} =  - \int  \ve{m}_{if} (\ve{q})\ {\rm d} \ve{P}_i \ .
$$
In case of reaction (1), one has 
$$
{\ve{D}}_{if}= i \omega ^{-1}\
\int_0^1 {\nabla}_{ \lambda \vsm{q} } [\ (\ \omega + M_{\rm d} - \sqrt{M_{\rm d}^2 +
(1-\lambda )^2 \ve{q}^2 } \ )\ J_{if}^0 (\lambda \ve{q})\ ]\ {\rm d}\lambda \ ,\ \eqno(13)
$$
$$
{\ve{M}}_{if} = i \int_0^1 {\nabla}_{\lambda \vsm{q} }\times {\ve{J}}_{if} (\lambda \ve{q}) 
\ \lambda \ {\rm d}\lambda \ ,\ \eqno(14) 
$$
where $J^{\nu}_{if} (\lambda \ve{q})$
is obtained from $\langle \pi ^+ {\rm nn;out} \mid J^{\nu}(0) \mid {\rm d}\ \rangle $
replacing the deuteron momentum by $(1-\lambda )\ve{q}$. 

It should be noted that, whereas quantities $\ve{d}_{if}$ and $\ve{m}_{if}$ in Eq.~(11) 
are singular and not proportional to the delta function expressing the momentum 
conservation, the representation (9) is free of singularities. Furthermore, it has been 
derived here (cf. Ref.~\cite{FriarFall}) without decomposition of the e.m. current into 
the part associated with the motion of the hadronic system as a whole and 
the intrinsic current and, therefore, can be employed in relativistic calculations. 

This representation generates a correction term additional to the ``canonical'' expression (3), 
which restores the GI of the amplitude in calculations that fail 
to satisfy the requirement (4). However, when the GI condition (4) does hold, 
this correction is equal to zero automatically.

In the long-wave limit, Eq.~(9) provides the fulfilment of the Siegert theorem~\cite{Siegert} 
for electric transitions in reactions with nonmeson channels~\cite{LS93}. For pion photoproduction 
on the free nucleon at threshold, it leads (as shown in Ref.~\cite{LS95}) to the Kroll-Ruderman 
result~\cite{KrollRud} emerging here as a particular case of the Siegert theorem. 
The results of the application of the approach outlined above to calculation 
of the differential cross sections of forward-angle pion electroproduction on the deuteron 
are presented in the next section. 

%\bigskip
%\centerline{\bf IV. RESULTS AND DISCUSSION}
%\bigskip
\section{Results and discussion}

In the Saclay experiment~\cite{Gilman}, determination has been made of the ratio $R$ of
the forward-angle $\pi ^+$ electroproduction cross section for the deuteron to that for 
the proton at two values (1160 MeV and 1230 MeV) of the (virtual) photon-nucleon 
invariant mass $W$. The incident beam energy $E$ and the electron-scattering angle 
were 645 MeV and $\theta = 36^{\circ }$, respectively, while pions were detected at 
$\ve{q}\hat{} \ve{p}_{\pi } = 4^{\circ }$. In particular, for $W=1160$ MeV ($\omega = 
290$ MeV, $\mid\ve{q}\mid =$414 MeV/c), the quantity 
$$
R = { 
{ \int [{\rm d}^4\sigma_{\rm d} / 
{\rm d}\Omega '{\rm d}E' {\rm d}\Omega_{\pi} {\rm d}E_{\pi} ]\ {\rm d}E_{\pi} }
\over 
{  {\rm d}^3\sigma_{\rm p} / 
{\rm d}\Omega '{\rm d}E' {\rm d}\Omega_{\pi} } 
}
\ =\ 0.80\pm 0.05 \eqno(15)
$$
was obtained, with the d$^4\sigma_{\rm d} $ deuteron cross section being integrated over 
the 50 MeV wide peak region. It was claimed in Ref.~\cite{Loucks} that the suppression of pion 
production rate for the deuteron compared to that for the proton is mostly due to 
the final-state interaction (FSI) between the two outgoing neutrons. However, the 
corresponding calculations were performed for point-like hadrons, while the neglect of 
the e.m. hadron structure is not quite adequate to the kinematical 
conditions of Ref.~\cite{Gilman}. Particularly, the calculated value of 153 pb/MeV$\cdot$sr$^2$ 
for the $^1{\rm H}({\rm e,e}'\pi ^+$)n cross section at $W=$1160 MeV turns out to be much 
larger than the measured quantity $46\pm 3$ pb/MeV$\cdot$sr$^2$. Furthermore, no 
comparison was made in Ref.~\cite{Loucks} between the plane-wave approximation results and 
those with the nn FSI included.

In the present paper, we apply the formalism outlined above to calculations of 
the pion production cross sections near threshold, in order to estimate the role 
of such ``nuclear medium'' effects as Fermi motion, binding, FSI of the nn pair, etc. in 
the quenching of the $\pi^+$ meson yield in the $^2{\rm H}({\rm e,e}'\pi ^+$)nn reaction 
observed in Ref.~\cite{Gilman}.

\begin{figure}[hbtp]
  \begin{center}
    \resizebox{13cm}{!}{\includegraphics{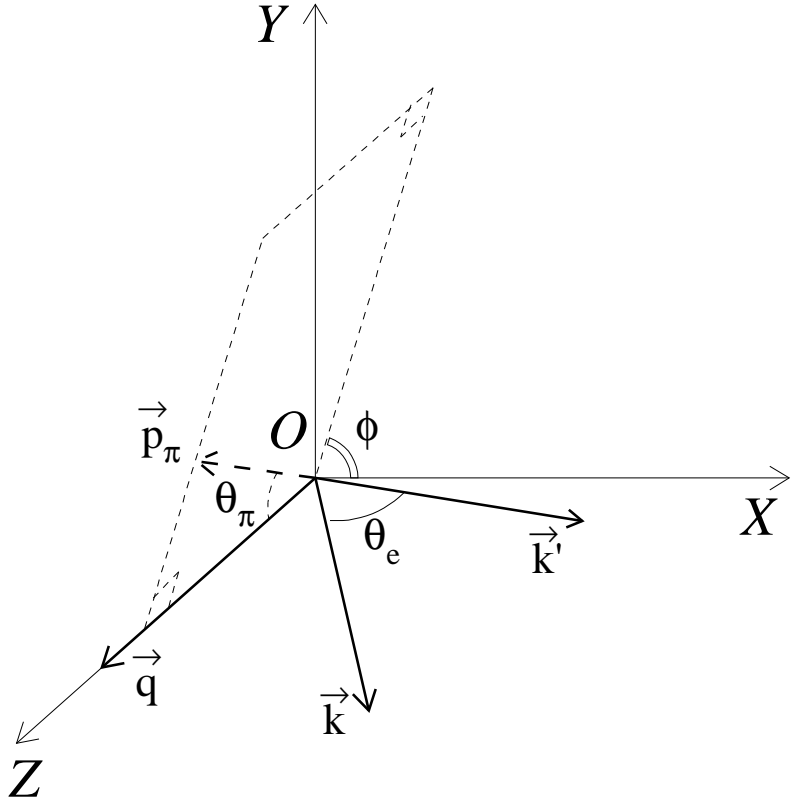}}
    \caption{Reference frame used in calculations.}
    \label{2}
  \end{center}
\end{figure}

The differential cross section of the $^2{\rm H}({\rm e,e}'\pi ^+$)nn reaction with 
unpolarized particles in the OPEA can be expressed~\cite{Forest} through the four SF's 
$W_{\alpha }$ ($\alpha =$C, T, I and S): 

$$
{ { {\rm d}^4 \sigma }\over {  {\rm d}\Omega '\ {\rm d}E'\ {\rm d}{\Omega}_
{\pi}\ {\rm d}E_{\pi} } }\ =
{\sigma}_{\rm M} E_{\pi}\mid \ve{p}_ {\pi}\mid\ ( \ \xi ^2 W_{\rm C}\ +\ 
(-{\xi \over{2}} + \eta ) W_{\rm T} \ +
$$
$$
+\  \xi ( -\xi + \eta ) ^{1\over 2}\cos \varphi W_{\rm I}\ -\ 
{{\xi }\over{2}}\cos 2\varphi \ W_{\rm S} \ ) \ ,\eqno(16)
$$
$$
\xi =q^2 / \ve{q}^2\ ,\ \ \ \eta = {\rm tan}^2 \theta /2 \ ,
$$
where $\sigma _{\rm M}$ is the Mott cross section, $q^{\mu }=(\omega,\ \ve{q})=(E-E',\ \ve{k} 
-\ve{k'}),\ q^2 = \omega ^2 -\ve{q}^2$,  and the SF's are related to the 
components of the hadronic tensor $W_{\mu \nu}$ by
$$
W_{\rm C}=W_{00}\ ,\ \ \ W_{\rm T}=W_{xx}+W_{yy}\ ,\ \ \ 
W_{\rm I}=W_{0x}+W_{x0}\ ,\ \ \ W_{\rm S}=W_{xx}-W_{yy}\ ,\eqno(17)
$$
$$
W_{\mu\nu} = (2\pi )^6 \ \int \hspace{-1.5 em} \sum_{\rm nn} \ \delta 
(q+P_{\rm d}-p_{\pi}-P_{\rm nn})\ F_{\rm \mu}^{*} F_{\rm \nu} 
\eqno(18)
$$
with
$$
F_{0} = i\ve{qD}_{if}\ ,
$$
$$
\ve{F} = i\omega \ve{D}_{if} - i\ve{q} \times \ve{M}_{if}\ .
$$
One may also write 
$$
F^{\nu} = J_{if}^{\nu} + G^{\nu}\ ,
\eqno(19)
$$
$$
J_{if}^{\nu} \equiv J_{if}^{\nu} (\ve{q}) \ ,
$$
where $G^{\nu}$ is the correction restoring the GI, as explained in Sec.~3, 
defined as 
$$
G^{0} = -{{ \omega + M_{\rm d} - \sqrt{M_{\rm d}^2 +
\ve{q}^2 } }\over {\omega }} J_{if}^0 (\lambda \ve{q} =0)\ ,
\eqno(20)
$$
$$
\ve{G} = \int_0^1 \nabla_{\lambda \ve{q}} \{ (\omega + M_{\rm d} - \sqrt{M_{\rm d}^2 +(1-\lambda )^2 
\ve{q}^2 }) J_{if}^0 (\lambda \ve{q}) - \lambda \ve{q} \ve{J}_{if} (\lambda \ve{q})  \}\ {\rm d} \lambda \ .
\eqno(21)
$$
Here we use the reference frame with the $OZ$ axis directed along $\ve{q}$, 
while the momenta of incident ($\ve{k}$) and scattered ($\ve{k'}$) electrons define 
the scattering plane $XOZ$ (see Fig.~\ref{2}).  
%%%%%%%%%%%%%%%%%%%%%%%% FIGURE 2 %%%%%%%%%%%%%%%%%%%%%%%%%%%%%%%%%%%%%%%%%%%%%% 
%\begin{figure}[h]
%\epsfxsize=12cm
%\epsffile{figure2.eps}
%%\caption{}
%
%{\small FIG. 2. Reference frame used in calculations..}
%\end{figure}
%%%%%%%%%%%%%%%%%%%%%%%%%%%%%%%%%%%%%%%%%%%%%%%%%%%%%%%%%%%%%%%%%%%%%%%%%%%%%%$$
%\ve{e}_Z = { {\ve{q}}\over {\mid\ve{q}\mid} }\ ,
%\ \ \ \ve{e}_Y = { {\ve{k}\times\ve{k'}}\over {\mid\ve{k}\times\ve{k'}\mid} }\ ,
%\ \ \ \ve{e}_X = \ve{e}_Y \times \ve{e}_Z \ .
%$$
The SF's in Eq.~(16) are taken for the angle $\varphi =0$ between vectors 
$\ve{q}\times\ve{p}_{\pi}$ and $\ve{e}_Y$. Averaging over target 
polarization states in formula (17) is implied.

\begin{figure}[hbtp]
  \begin{center}
    \resizebox{12cm}{!}{\includegraphics{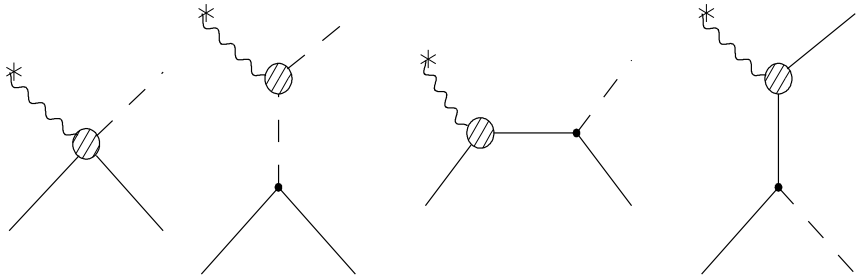}}
    \caption{Feynman graphs corresponding to ${\gamma}^* {\rm N} \rightarrow \pi {\rm N}'$ 
reaction amplitude in the generalized Born approximation based on the PV $\pi$NN coupling 
given by Ref.~\protect\cite{DombeyRead}.}
    \label{3}
  \end{center}
\end{figure}

To construct the operator of pion production on a bound nucleon, we have exploited, 
as an example, the so-called generalized Born approximation based on the pseudovector 
(PV) $\pi$NN coupling (see, e.g., Ref.~\cite{DombeyRead}) with the e.m. vertices taken for the 
on-mass-shell hadrons. The corresponding Feynman graphs are shown in Fig.~\ref{3}. 
%%%%%%%%%%%%%%%%%%%%%%%% FIGURE 3 %%%%%%%%%%%%%%%%%%%%%%%%%%%%%%%%%%%%%%%%%%%%%% 
%\begin{figure}[h]
%\epsfxsize=12cm
%\epsffile{figure3.eps}
%%\caption{}
%
%{\small FIG. 3. Feynman graphs corresponding to ${\gamma}^* {\rm N} \rightarrow \pi {\rm N}'$ reaction amplitude 
%in the generalized Born approximation based on the PV $\pi$NN coupling.}
%\end{figure}
%%%%%%%%%%%%%%%%%%%%%%%%%%%%%%%%%%%%%%%%%%%%%%%%%%%%%%%%%%%%%%%%%%%%%%%%%%%%%
Parametrizations~\cite{KornKur} for the nucleon e.m. form factors were 
used, while expressions for the nucleon axial and pion form factors were taken from 
Ref.~\cite{EW}. Unlike calculations~\cite{Loucks}, the Fermi-motion effects have been included 
in the operator without any nonrelativistic reduction. 
The FSI of the pion as well as the $\Delta_{33}$ resonance 
production were neglected in present calculations.

Using this operator and the explicitly gauge independent expression (9) for the 
amplitude, we have calculated the differential cross sections of $^1{\rm H}({\rm e,e}'\pi ^+$)n 
and $^2{\rm H}({\rm e,e}'\pi ^+$)nn reactions for the kinematics of Ref.~\cite{Gilman} at $W=$1160 MeV.

For d$^3\sigma_{\rm p}/{\rm d}\Omega '{\rm d}E' {\rm d}\Omega_{\pi}$, the value of 
65.3 pb/MeV$\cdot$sr$^2$ has been obtained. It turns out also that 
making the gauge subtraction (8) in the 
amplitude instead of application 
%%%%%%%%%%%%%%%%%%%%%%%% FIGURE 4 %%%%%%%%%%%%%%%%%%%%%%%%%%%%%%%%%%%%%%%%%%%%%% 
%\begin{figure}[h]
%\epsfig{figure=figure4.eps,width=12cm}
%
%{\small FIG. 4. Longitudinal and transverse cross sections of the forward-angle 
%pion electroproduction on the proton in the c.m. frame. The solid and dashed 
%curves display results obtained using the extension of the Siegert theorem and 
%the subtraction procedure, respectively. The experimental data are from Ref.~\protect\cite{Bardin}.}
%\end{figure}
%%%%%%%%%%%%%%%%%%%%%%%%%%%%%%%%%%%%%%%%%%%%%%%%%%%%%%%%%%%%%%%%%%%%%%%%%%%%%%%% 
\begin{figure}[hbtp]
  \begin{center}
    \resizebox{12cm}{!}{\includegraphics{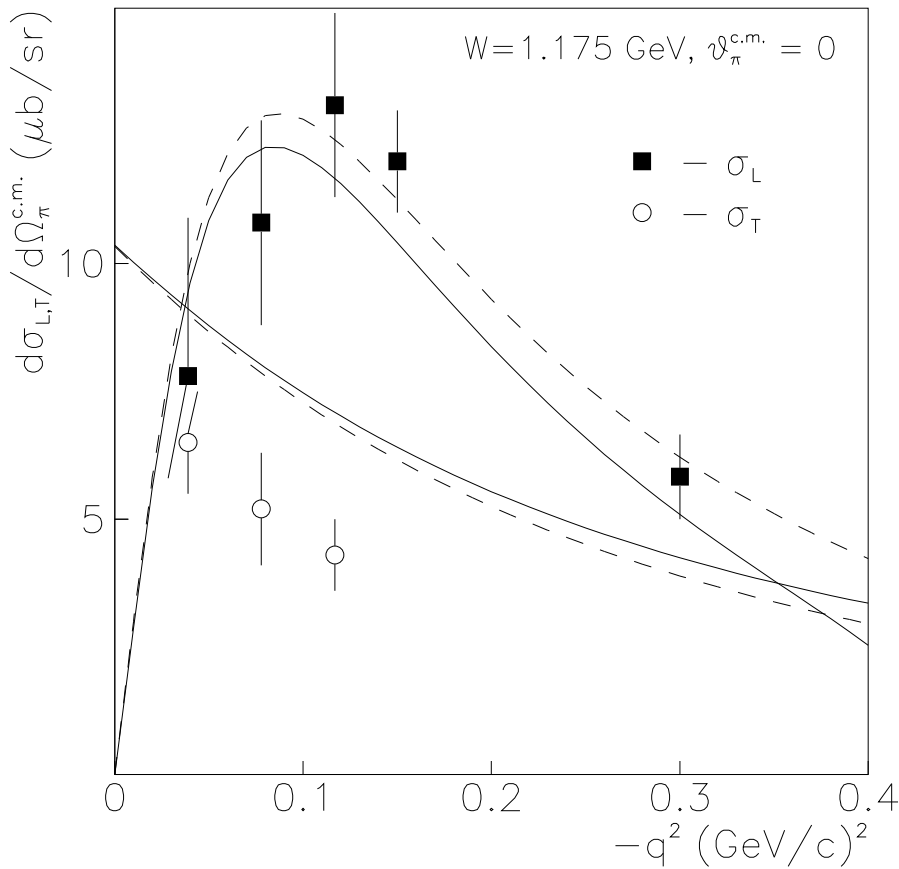}}
    \caption{Longitudinal and transverse cross sections of the forward-angle 
pion electroproduction on the proton in the c.m. frame. The dashed and solid 
curves display results obtained with subtraction (8) and representation (9) for the amplitude, 
respectively. The experimental data are from Ref.~\protect\cite{Bardin}.}
    \label{4}
  \end{center}
\end{figure}
of Eq.~(9) leads to the same result, which is some 
greater than the observed quantity. To clarify the origin of this discrepancy, we have 
calculated the c.m. longitudinal and transverse cross sections of 
$\pi^+$ meson electroproduction on the proton for the kinematics of experiment~\cite{Bardin} 
(see \hbox{Fig.~\ref{4}}). 
The results for the longitudinal cross section are in fair agreement with data. 
The description of the transverse cross section as well as our results for the 
kinematics~\cite{Gilman} could be improved by taking into account the $\Delta_{33}$ resonance 
contribution to the amplitude, which would affect the transverse component of the pion 
production operator with practically no changes on the longitudinal one. 

When calculating the differential cross sections for the deuteron, we made use of the 
parametrizations~\cite{Paris1} for the $S$- and $D$-components of the deuteron wave function. 
The nn FSI in the $^1 S_0$ state, which gives a predominant contribution at very 
small relative momenta of the outgoing neutrons, was calculated for the Paris 
potential~\cite{Paris2} exploiting version~\cite{KorSheb} of the matrix inversion method~\cite{BJ}. 
The results obtained within the approach outlined in 
%%%%%%%%%%%%%%%%%%%%%%%% FIGURE 5 %%%%%%%%%%%%%%%%%%%%%%%%%%%%%%%%%%%%%%%%%%%%%% 
%%\begin{wrapfigure}[25]{r}[0pt]{16cm}
%%\epsfig{figure=figure5.eps,width=9cm}
%%{\small Fig. 5. Differential cross sections of the $^2{\rm H}(\gamma, \pi ^+$)nn 
%%reaction for $\omega =170$ MeV,$ \theta_{\pi}=36^{\circ}$ at low energy of the nn pair. 
%%The solid line shows the result obtained within our approach, while the dashed 
%%(dotted) curve refers to the IA calculations with (without) the $^1 S_0$-state 
%%nn FSI included. The histogram exhibits data \protect\cite{Mainz}.}
%%\end{wrapfigure}
%\begin{figure}[h]
%\epsfxsize=12cm
%\epsffile{figure5.eps}
%%\caption{}
%
%{\small FIG. 5. Differential cross sections of the $^2{\rm H}(\gamma, \pi ^+$)nn 
%reaction for $\omega =170$ MeV, $ \theta_{\pi}=36^{\circ}$ at low energy of the nn pair. 
%The solid line shows the result obtained within our approach, while the dashed 
%(dotted) curve refers to the IA calculations with (without) the $^1 S_0$-state 
%nn FSI included. The histogram exhibits data \protect\cite{Mainz}.}
%\end{figure}
%%%%%%%%%%%%%%%%%%%%%%%%%%%%%%%%%%%%%%%%%%%%%%%%%%%%%%%%%%%%%%%%%%%%%%%%%%%%%%%% 
\begin{figure}[hbtp]
  \begin{center}
    \resizebox{12cm}{!}{\includegraphics{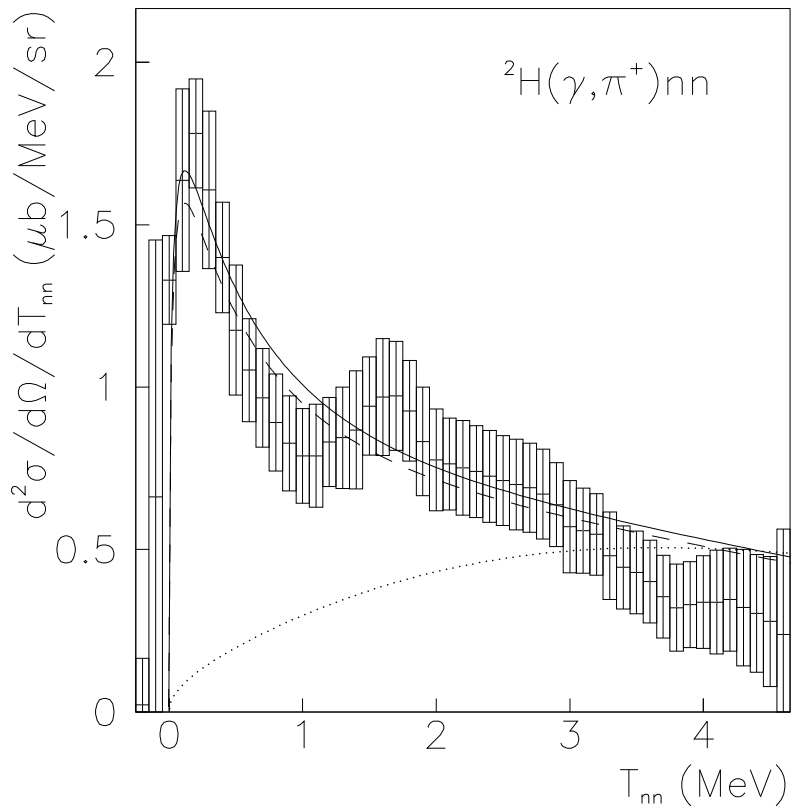}}
    \caption{Differential cross sections of the $^2{\rm H}(\gamma, \pi ^+$)nn 
reaction for $\omega =170$ MeV, $ \theta_{\pi}=36^{\circ}$ at low energy of the nn pair. 
The solid line shows the result obtained within our approach, while the dashed 
(dotted) curve refers to the IA calculations with (without) the $^1 S_0$-state 
nn FSI included. The histogram exhibits data \protect\cite{Mainz}.}
    \label{5}
  \end{center}
\end{figure}
Secs.~2 and 3 were compared 
with those given by the IA. In the latter case, the on-shell extrapolation for the 
one-body contribution to the amplitude has been chosen in a way that provides the 
fulfilment of the GI requirement (4) at the photon point $q^2=0$, while, for 
$q^2 \neq 0$, subtraction (8) was made.

In order to estimate the quality of our treatment of the nn FSI, we applied  
these approaches to reproduce Mainz data~\cite{Mainz} on $\pi^+$ meson photoproduction 
from the deuteron at small energies $T_{\rm nn}$ of the residual nn system (see 
Fig.~\ref{5}). It is seen that both the approaches give results in reasonable 
agreement with the experiment. It turns out also that the IA curve practically coincides 
with the calculations presented in Ref.~\cite{Mainz}.

\begin{figure}[hbtp]
  \begin{center}
    \resizebox{11cm}{!}{\includegraphics{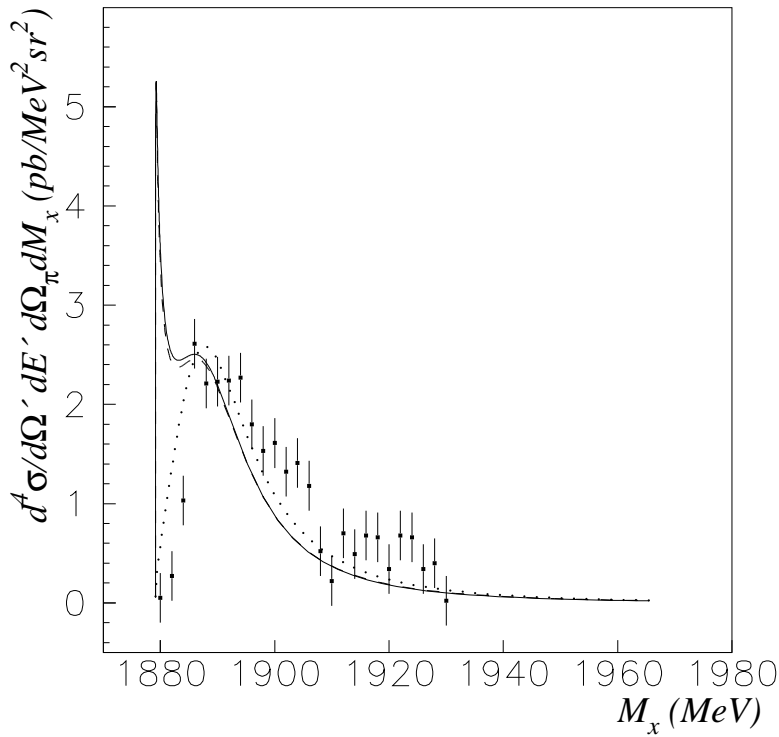}}
    \caption{Differential cross sections of the $^2{\rm H}({\rm e,e}'\pi ^+$)nn 
reaction for the kinematics of experiment~\protect\cite{Gilman}. Notations for the curves are the 
same, as in Fig.~\protect\ref{5}.}
    \label{6}
  \end{center}
\end{figure}

The plots of the $^2{\rm H}({\rm e,e}'\pi ^+$)nn cross section obtained for the kinematics 
of the Saclay experiment~\cite{Gilman} are displayed in Fig.~\ref{6}. 
The cross section is presented as a function of the ``missing'' mass $M_x$ (invariant mass 
of the nn pair). The shown results suggest 
that, contrary to the conclusions of Ref.~\cite{Loucks}, the nn FSI leads to some increase 
%%%%%%%%%%%%%%%%%%%%%%%% FIGURE 6 %%%%%%%%%%%%%%%%%%%%%%%%%%%%%%%%%%%%%%%%%%%%%% 
%%\epsfig{figure=figure6.eps,width=9cm}
%%{\small Fig. 6. Differential cross sections of the $^2{\rm H}({\rm e,e}'\pi ^+$)nn 
%%reaction for the kinematics of experiment \protect\cite{Gilman}. The notation for the curves is the 
%%same, as in Fig. 3.}
%%\end{wrapfigure}
%\begin{figure}[ht]
%\epsfxsize=12cm
%\epsffile{figure6.eps}
%%\caption{}
%
%{\small FIG. 6. Differential cross sections of the $^2{\rm H}({\rm e,e}'\pi ^+$)nn 
%reaction for the kinematics of experiment \protect\cite{Gilman}. Notations for the curves are the 
%same, as in Fig. 5.}
%\end{figure}
%%%%%%%%%%%%%%%%%%%%%%%%%%%%%%%%%%%%%%%%%%%%%%%%%%%%%%%%%%%%%%%%%%%%%%%%%%%%%%%% 
of the $\pi^+$ production rate within the pion energy range covered in the experiment. 
Indeed, the ratio $R$ (see Eq.~(15)) calculated in the plane-wave IA is 0.74, whereas 
the corresponding value given by the IA with the $^1 S_0$ nn FSI included turns out 
to be 0.79. It is seen also that the result obtained using the unitary transformation 
method in conjunction with the extension of the Siegert theorem does not much differ 
from the one provided by the IA indicating a weak sensitivity 
of the $^2{\rm H}({\rm e,e}'\pi ^+$)nn differential cross section to the two-body 
contributions for the kinematics of Ref.~\cite{Gilman}. In this case, $R=0.80$ was derived 
being in agreement with the measured quantity (15).

Thus, our calculations suggest that the main source of the $\pi^+$ meson yield 
quenching observed in the experiment~\cite{Gilman} in case of the deuteron target originates 
from nuclear medium effects influencing the $\pi^+$ production process. These effects 
include nucleon Fermi motion and binding along with Pauli blocking of the final nn state.   
The Fermi-motion and binding effects are involved in the one-body pion production operator 
as well as in the phase-space factor arising due to the energy-momentum conservation law, 
when performing integration in Eq.~(18). 
The binding effects in the operator emerge as an off-shell behavior of the one-body amplitude. 
Their importance in description of exclusive pion production off the deuteron was pointed out 
in Ref.~\cite{LS99}. 
However, the corresponding contribution to the 
$^2{\rm H}({\rm e,e}'\pi ^+$)nn cross section for the kinematics of Ref.~\cite{Gilman} 
appears to be not significant. 

\begin{figure}[hbtp]
  \begin{center}
    \resizebox{12cm}{!}{\includegraphics{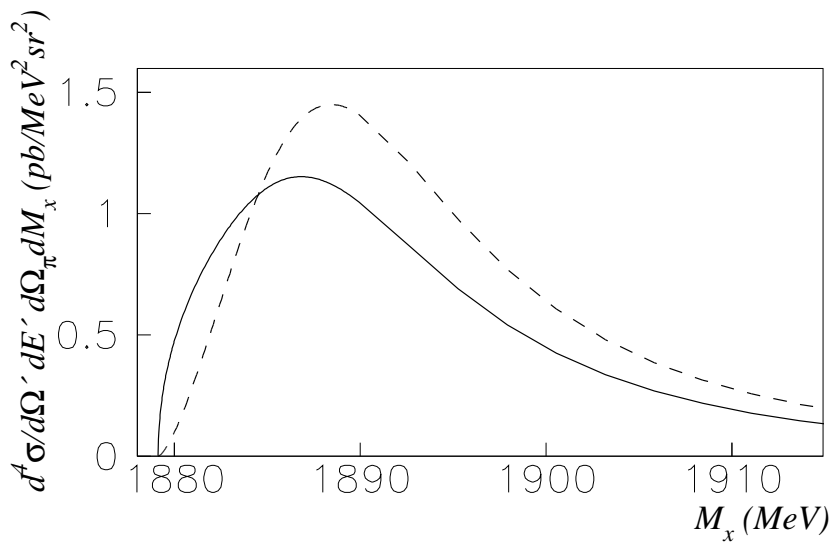}}
    \caption{Contributions of the singlet (solid line) and triplet (dashed curve) final nn states 
    in the $^2{\rm H}({\rm e,e}'\pi ^+$)nn differential cross section 
calculated within the plane-wave IA  for the kinematics of experiment \protect\cite{Gilman}.}
    \label{7}
  \end{center}
\end{figure}

The other factors of nuclear medium influence mentioned above 
prove to be important and result in the overall pion production rate reduction by 
more than 20\%. 
In particular, it turns out that the Pauli suppression in the final two-neutron state 
reduces the ratio (15) by about the same value, as the nn FSI increases it. 
The role of this effect is also illustrated by Fig.~\ref{7}, where the contributions 
due to the singlet and triplet final nn states to the differential cross section 
calculated within the plane-wave IA are displayed. 
%%%%%%%%%%%%%%%%%%%%%%%% FIGURE 7 %%%%%%%%%%%%%%%%%%%%%%%%%%%%%%%%%%%%%%%%%%%%%% 
%%%%%%%%%%%%%%%%%%%%%%%%%%%%%%%%%%%%%%%%%%%%%%%%%%%%%%%%%%%%%%%%%%%%%%%%%%%%%%%% 
It is seen that the Pauli antisymmetrization of the two-nucleon wave function 
strongly suppresses the triplet contribution in the region of small relative momenta 
of the outgoing neutrons. 
It should be noted also that the incompleteness of the pion energy range coverage in the integral 
of Eq.~(15) also contributes to reduction of the ratio given by that formula. 

For a quasiparallel kinematics and a relatively small momentum transferred to the hadronic system 
(as it takes place in the experiment~\cite{Gilman}), the transverse components $T_{x,y}^{[1]}$ 
of the pion production operator are dominated by the spin-flip Kroll-Ruderman 
contribution~\cite{KrollRud}, whereas the ``longitudinal'' virtual photons couple mostly 
to a term proportional to the component $\sigma_z$ of the Pauli spin matrix $\ve{\sigma}$ 
giving no rise to spin-flip transitions. Therefore, the singlet nn final state contribution, e.g., 
in the purely transverse SF's 
$W_{\rm T}$ and $W_{\rm S}$ results mainly from the deuteron states with spin projections 
$M_{\rm d} = \pm 1$, while the longitudinal hadronic response in the $^1 S_0$ nn FSI peak 
region is determined by the $M_{\rm d} = 0$ deuteron state. This consideration complies with 
the conclusions of Ref.~\cite{Loucks} indicating a strong $\pi^+$ production rate dependence 
on the deuteron spin orientation in the vicinity of the $^1 S_0$ peak which, in principle, 
could be observed experimentally with a polarized deuteron target. 
As pointed out in paper~\cite{Loucks}, this result is predetermined 
by the angular momentum conservation law in the reaction 
in case of an exactly parallel kinematics.   
We see, that the spin correlations discussed here should take place also for moderate 
deviations from the parallel geometry. However, they might be veiled then 
by the pion FSI and by extra terms in the pion production operator for kinematics 
probing high-momentum components of the deuteron wave function.   

When comparing our numerical results with experiment~\cite{Gilman}, it should be kept in mind 
that the data plotted in Fig.~\ref{6} have not been corrected for radiative processes. Introduction of 
such corrections could shift the experimental points by several MeV towards lower 
missing masses making the agreement with our calculations more fair. 
Radiation broadening along with finite spectrometer resolution 
might also be the reason of the fact that the FSI peak was not clearly resolved in the experiment. 

\begin{figure}[hbtp]
  \begin{center}
    \resizebox{12cm}{!}{\includegraphics{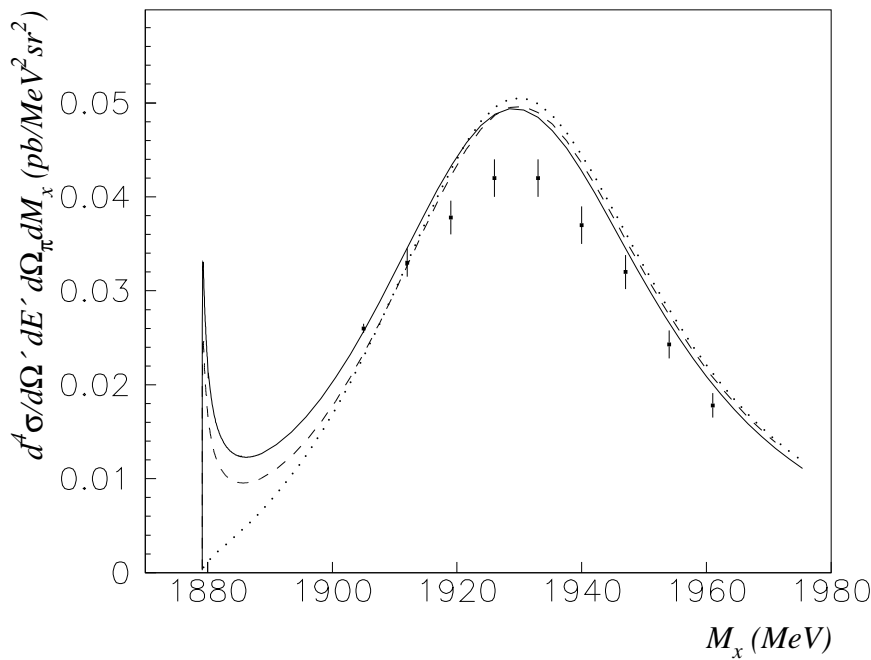}}
    \caption{Differential cross sections of the $^2{\rm H}({\rm e,e}'\pi ^+$)nn 
reaction for the kinematics of experiment \protect\cite{Gaskell}. Notations for the curves are the 
same, as in Figs.~\protect\ref{5} and \protect\ref{6}.}
    \label{8}
  \end{center}
\end{figure}

Recent measurements of cross sections of forward-angle $\pi^+$ 
electroproduction on the lightest nuclei with the physics motivation similar to that of 
paper~\cite{Gilman} have been carried out in Jefferson Lab experiment E91003~\cite{Gaskell}. 
Our calculations 
for the $^2{\rm H}({\rm e,e}'\pi ^+$)nn reaction at one of the E91003 kinematical sets 
($q^2$ =-0.4 (GeV/c)$^2$, $W$=1160 MeV, $E$=845 MeV, and $\omega$= 450 MeV) 
are presented in Fig.~\ref{8}. Near the quasi-free peak, our approach leads to the results almost 
indistinguishable from the ones given by the IA. However, we observe an appreciable discrepancy 
for lower missing 
masses (greater pion energies) that reaches 20 \% in the $^1 S_0$ nn FSI peak region. 
In general, the results of calculations are in a satisfactory agreement with the experiment, 
though in the vicinity of the quasi-free peak, the curves are about 15 \% above the data. 
It could be explained by a more considerable relative contribution of the transverse SF $W_{\rm T}$ to the cross section 
for the selected kinematics compared to kinematics of Ref.~\cite{Gilman}. (The ratio $(-{\xi \over{2}} + \eta ) / \xi ^2$ 
[see Eq.~(16)] is equal to 1.7 and 1.4, respectively). Since this SF is sensitive to the $\Delta _{33}$ 
resonance production mechanism neglected in the present calculations, it could result in some overestimate 
of the cross section (cf. Fig.~\ref{4}). As shown in Ref.~\cite{GanSheb}, for kinematics tuned to the $\Delta _{33}$ resonance, 
a considerable cancellation occurs between the resonance contributions to the one-body amplitude 
and the pion rescattering terms. However, for the considered above kinematic conditions which are away 
from the $\Delta _{33}$ peak for the system of the virtual photon and the proton, the $\Delta _{33}$ contribution 
to the cross section can be still significant through the interference with the nonresonance terms displayed in 
Fig.~\ref{3}. So, to make a more definite conclusion about the true role of the $\Delta _{33}$ resonance production, 
a thorough study of FSI including 3-body dynamics is needed (see, e.g., Ref.~\cite{Luciano}). 
We would like to note also that the choice of kinematics for experiments~\cite{Gilman,Gaskell} was predetermined by 
their physics motivation with a hope to have the cross sections strongly dominated by the pion-pole 
pion production mechanism. It turns out, however, that other processes shown in Fig.~\ref{3} also play an essential 
role, in particular, via their interference with the pion-pole diagram. In this connection, our calculations 
confirm the conclusions of Refs.~\cite{Loucks,HafLee}. 
%%%%%%%%%%%%%%%%%%%%%%%% FIGURE 8 %%%%%%%%%%%%%%%%%%%%%%%%%%%%%%%%%%%%%%%%%%%%%% 
%%\epsfig{figure=figure8.eps,width=9cm}
%%{\small Fig. 8. Differential cross sections of the $^2{\rm H}({\rm e,e}'\pi ^+$)nn 
%%reaction for the kinematics of experiment \protect\cite{Gilman}. The notation for the curves is the 
%%same, as in Fig. 5.}
%%\end{wrapfigure}
%\begin{figure}[ht]
%\epsfxsize=12cm
%\epsffile{figure7.eps}
%%\caption{}
%
%{\small FIG. 8. Differential cross sections of the $^2{\rm H}({\rm e,e}'\pi ^+$)nn 
%reaction for the kinematics of experiment \protect\cite{Gaskell}. Notations for the curves are the 
%same, as in Figs. 5 and 6.}
%\end{figure}
%%%%%%%%%%%%%%%%%%%%%%%%%%%%%%%%%%%%%%%%%%%%%%%%%%%%%%%%%%%%%%%%%%%%%%%%%%%%%%%% 

%\bigskip
%\centerline{\bf V. CONCLUSION}
%\bigskip
\section{Conclusion}

The approach based 
on the unitary transformation method has been applied to the treatment of pion 
electroproduction on the deuteron near threshold. To provide the GI of the results, 
we have used the expression for the amplitude stemming from the extension of the 
Siegert theorem. Within the approach, the differential cross section of the 
$^2{\rm H}({\rm e,e}'\pi ^+$)nn reaction for the kinematics of the Saclay experiment has 
been calculated and compared with that given by the IA. The results suggest that the 
observed suppression of the $\pi^+$ meson production rate in the $^2{\rm H}({\rm e,e}'\pi ^+$)nn 
reaction compared to the case of pion electroproduction on the proton can be due to 
such effects as nucleon Fermi motion and binding. Calculations performed 
for the kinematics of Jefferson Lab experiment E91003 also manifest a satisfactory 
agreement with the data. Further improvement of the model (in particular, its extension 
to the $\Delta_{33}$ resonance region) can be achieved by taking into account the FSI of 
the pion with the nn system.

\bigskip

This work was supported in part by the grant awarded by the Deutsche Bundesministeriums f\"ur 
Bildung, Wissenschaft, Forschung und Technologie. LL thanks the University of Padova 
and INFN for support and hospitality in October~-- November, 2003.

%\newpage
%\centerline{\bf References}

%\vspace{0.2cm}
%\vfill

\end{document}